# Efficient Asynchronous RPC Calls for Microservices: DeathStarBench Study

Stijn Eyerman                                  Ibrahim Hur

stijn.eyerman@intel.com              ibrahim.hur@intel.com

Intel Corporation

## Introduction

Microservices is a popular paradigm for implementing interactive cloud applications (e.g., a social network) [1]. The application is split up into small functions, called services. Each service runs in its own process or container and communicates with other services through remote procedure calls (RPC). The main advantages of microservices are maintainability and scalability. If a function needs to be updated, only the corresponding service needs to be replaced, the rest of the application is untouched and keeps running. When one or more functions need to scale up because of increasing request rate, one just needs to increase the number of instances and/or the thread count of these services, again without impacting the rest of the application. Because RPC also works between separate machines, the applications can also be easily scaled up from one to multiple machines.

Crucial in the performance of microservice applications is the efficient handling of RPC calls. We found that the asynchronous call implementation in a popular microservice benchmark suite, DeathStarBench [2], causes a bottleneck in thread management, reducing the peak throughput and increasing latency at high request rates. Replacing the threaded implementation with a fiber implementation increases peak throughput by up to 6x, meaning that the service can operate at a higher request rate without significantly increasing response latency.

## Asynchronous RPC calls

A typical microservice is called by the client or by another microservice, and on its turn calls other microservices to collect data (e.g., ask a user id based on a username) or to continue the request from the client (e.g., inserting data into a database). Calling another microservice is done through asynchronous calls: the call is initiated without blocking the main thread. This means other instructions, including other RPC calls, can be done while the call is being serviced. A blocking 'get' function returns when the data is available. This way, multiple RPC calls can be done in parallel, overlapping their waiting times.

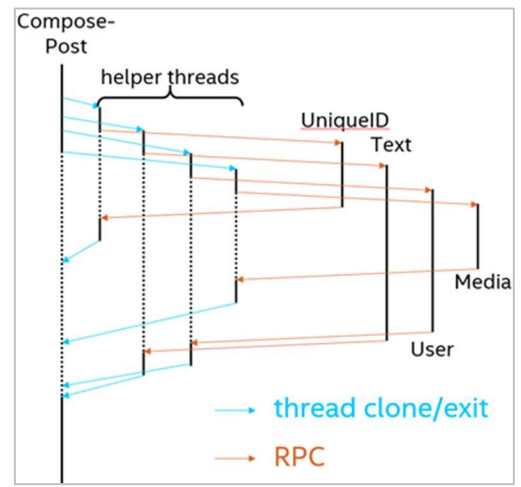

In DeathStarBench, these calls are implemented using the std::async library in C++. In its default implementation, it spawns a new thread for each async call, which then performs the RPC call to the destination service. This is illustrated in the figure at the right: the ComposePost service needs to call four other services (UniqueID, Text, User and Media). Thereto, it creates four threads to perform the RPC calls, and the main thread waits until all four threads have finished.

In this implementation, the RPC calls and waiting time overlap, exploiting parallelism to reduce latency. However, the creation and exiting of threads puts a high burden on the OS kernel. Simulations of the SocialNetwork application in DeathStarBench show that the ComposePost service spends 23% of its time in clone and exit system calls. When highly loaded, thread creation, scheduling and exiting can limit the throughput of the service, as thread management involves synchronization and serialization.

## Boost fibers

Boost [3] is a set of high-performance libraries for C++. Fibers were introduced in Boost version 1.62 (September 2016). Fibers are essentially microthreads, managed and executed within a single hardware thread. When a fiber (or multiple fibers) is spawned, the fiber scheduler is activated, which decides what to run: the main thread or one of the fibers. Only one is running at a time, because it runs in a single hardware thread. The advantage of fibers is that fiber management is much lighter than thread management (avoiding kernel calls). The main disadvantage is that fibers are not executed in parallel and do not exploit the performance potential of multiple cores. However, if a fiber is in a waiting state, it is scheduled out and replaced by another fiber, effectively overlapping waiting times. This is particularly interesting for asynchronous RPC calls, where the helper threads spend the majority of their time waiting for the RPC calls to finish. A fiber execution for the ComposePost helper threads is illustrated at the right. Note that only one fiber is running at a time. Nevertheless, the waiting time of the ComposePost main thread is not increased (much) compared to the threaded implementation.

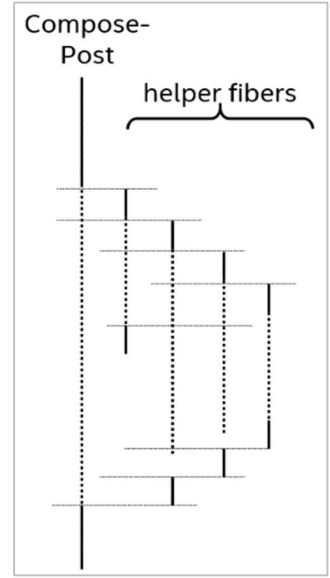

Fibers also have an asynchronous interface (boost::fiber::async) with the same signature as the std::async interface. Altering a microservice implementation to use fibers instead of threads involves installing a recent version of the boost library, including the fiber header files, and 'search-and-replace' the std::async calls by boost::fiber::async calls. Furthermore, the flexibility of microservices enables a smooth transition to the fiber implementation, by replacing the affected services one by one, without interrupting the service.

## Evaluation

We compare the default DeathStarBench implementation of SocialNetwork with an implementation where all asynchronous calls are replaced by fibers. To evaluate performance, we run the SocialNetwork microservices on a single Intel® Xeon® Platinum 8368 machine (Icelake architecture, 2 sockets, 38 cores per socket, 2 threads per core). We load the social network with 4 different request generators, provided by the benchmark developers: ComposePost, ReadUserTimeline, ReadHomeTimeline and mixed (a combination of these three requests).

We measure peak throughput and tail latency at different request rates. For peak throughput, we increase the request rate of the request generator until the number of processed requests per second does not increase anymore. This request rate is not a realistic operating point, because the latencies of the requests are very high. Therefore, we also measure tail latency at different request rates up to just beneath the peak throughput.

The first figure shows the peak throughput in requests per second. The peak throughput for the fiber implementation is higher than the thread implementation, especially for the compose (6x) and the mixed workload (3.6x). The ComposePost service has the largest number of async calls, which explains the high gain. Mixed also has a fraction of compose requests. Although peak throughput cannot be reached in reality, it is an indication of what throughput a service can sustain with reasonable latency.

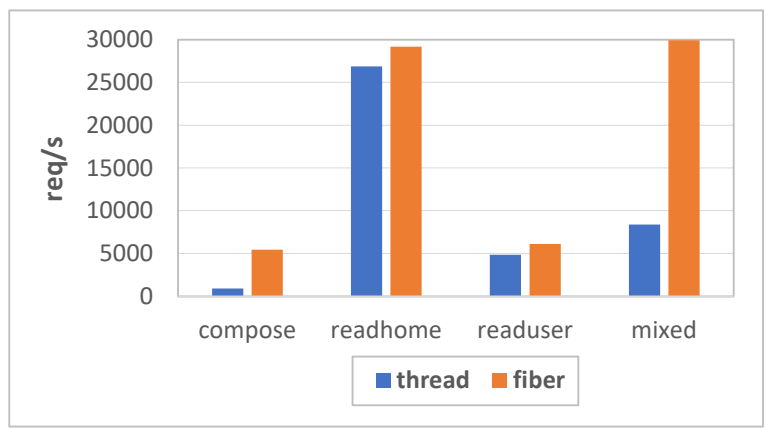

The figures beneath show the 99% tail latency (99[th] percentile of the latency distribution) for the four workloads as a function of request rate. The request rate is increased from a low value up to a rate close to the peak throughput of the threaded implementation. Beyond that rate, latency keeps increasing asymptotically. At low request rates, the tail latency of the fiber implementation is comparable to that of the thread implementation. It is sometimes a bit higher, because of the lower parallelism and thread pressure is low at that point. As the request rate approaches the peak throughput, the tail latency of the thread implementation quickly increases, while the fiber implementation has no significant increase in latency. Further increasing the request rate up to the fiber implementation peak throughput would also cause an exponential increase in the fiber implementation latency, but for the same request rate, the fiber implementation has a clearly lower latency.

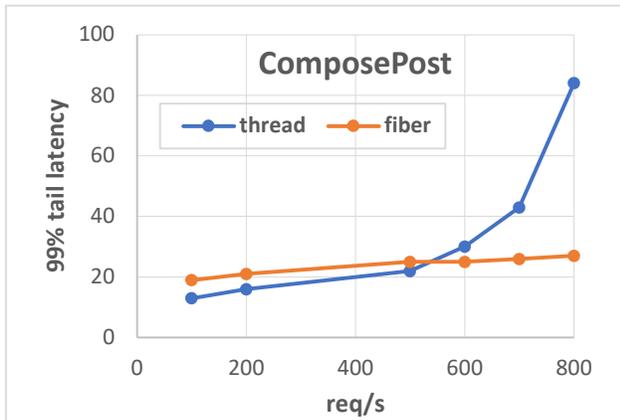
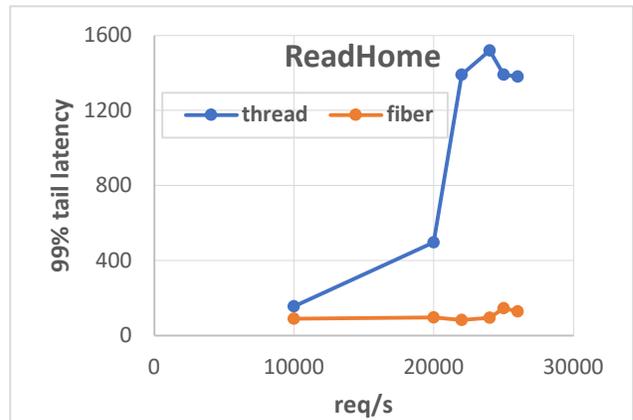
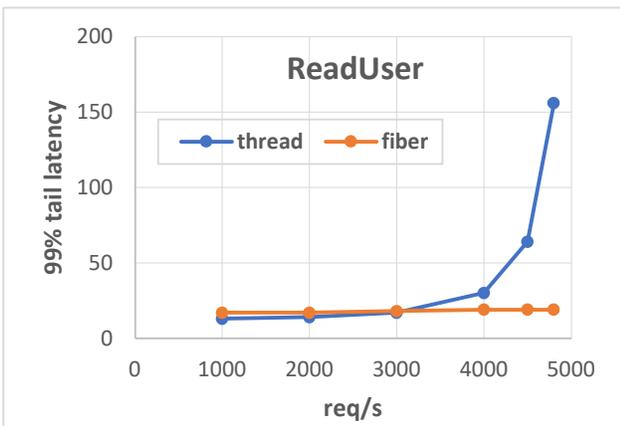
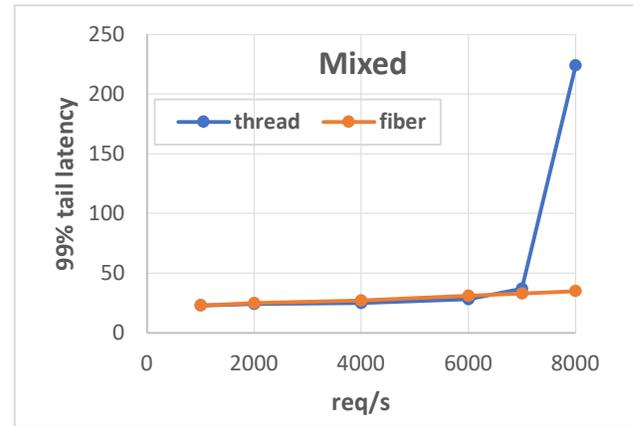

## Conclusions

Asynchronous calls are crucial for the performance of microservice applications because they enable executing microservices in parallel. Asynchronous calls are typically implemented using separate threads. However, if the request rate is high, thread management limits the peak throughput and increases tail latency. Boost fibers is a recent addition to the boost library, implementing lightweight micro-threads. Using fibers instead of threads significantly increases the peak throughput and reduces latency at high request rates. The increased peak throughput enables operating the service at a higher throughput with no increase in latency and without scaling the hardware.